\documentclass[10pt,a4paper]{article}

\usepackage{graphicx, amssymb}

\usepackage[centertags]{amsmath}

\normalsize \textwidth 454pt \textheight 690pt \baselineskip 14.5pt \columnsep 12pt \columnwidth
230pt \topskip 12pt \evensidemargin 26pt \oddsidemargin 9pt \topmargin -5 mm

\begin{document}

\title{On the extended Kolmogorov-Nagumo information-entropy theory, the $q \rightarrow \frac{1}{q}$ duality and its possible implications for a non-extensive two dimensional Ising model.}

\author{Marco Masi\footnote{Corresponding author: marco\_masi2@tin.it, fax: +39 02 6596997} \\ \small via Tiepolo 36, 35100, Padova, Italy.}

\maketitle

\begin{abstract}
The aim of this paper is to investigate the $q \rightarrow \frac{1}{q}$ duality in an
information-entropy theory of all q-generalized entropy functionals (Tsallis, Renyi and
Sharma-Mittal measures) in the light of a representation based on generalized exponential and
logarithm functions subjected to Kolmogorov's and Nagumo's averaging. We show that it is precisely
in this representation that the form invariance of all entropy functionals is maintained under the
action of this duality. The generalized partition function also results to be a scalar invariant
under the $q \rightarrow \frac{1}{q}$ transformation which can be interpreted as a non-extensive
two dimensional Ising model duality between systems governed by two different power law long-range
interactions and temperatures. This does not hold only for Tsallis statistics, but is a
characteristic feature of all stationary distributions described by q-exponential Boltzmann
factors.
\end{abstract}



\vspace{20mm}

To be published on Physica A

\newpage

\section{Introduction}

Because of a growing interest in non-extensive thermostatistics, during recent years, a good deal
of study has been devoted to a non-extensive generalization of Boltzmann-Gibbs statistical
mechanics.\cite{Tsallis2} A generalization of the ordinary formalism was however mainly devoted to
the study of C. Tsallis' entropy measure \cite{Tsallis} and separately to a lesser degree to
Renyi's entropy functional \cite{Renyi}, both indexed by a $q$-parameter. Tsallis measure has been
preferred over other q-generalized measures since the stability and concavity properties of
Renyi's entropy are still a matter of debate. Therefore only few efforts have been pursued to
unite them in a common and natural theoretical framework.

Nevertheless some attempts in this direction have been made, e.g. by Frank and Plastino
\cite{Frank} and Frank and Daffertshofer \cite{Daffertshofer}. In a previous communication
\cite{Masi}, a more analytic approach was used using a Kolmogorv-Nagumo generalized
logarithmic-exponential representation (KNG-representation). We could write down the set of
entropy functionals, from Boltzmann-Gibbs (BG) entropy through Renyi and Tsallis, up to
Sharma-Mittal (SM) \cite{Sharma} and a new type of entropy measure we called "supra-extensive".
Here the increasing generalization of entropy measures from arithmetic to non-arithmetic averages,
and from additive to non-additive systems became clear in its hierarchical structure (for similar
attempts see also \cite{Naudts}). With this formalism it is also easier to highlight the
relationship that exists between generalized information-entropy measures and Fisher
information.\cite{Masi2} It is so far uncertain if this can have a wider application in
thermostatistics, but since this representation unites in one frame all pseudoadditive and
KN-averaged measures leading to maximum entropy probability distributions (MEPDs) (of entropies
subjected to escort-means) exhibiting q-generalized Boltzmann factors, it is worthwhile to study
its overall structure.

In this paper we will investigate this deformed logarithm-exponential representation focussing our
attention on the dual description of generalized statistics. We show how in this representation a
deeper meaning of the $q \rightarrow \frac{1}{q}$ duality emerges, first studying to what kind of
formal invariance it leads and then highlighting in what way one can relate it to a non-extensive
Ising model duality.

\section{Overview of the generalizing path for entropy measures}

We shortly recall how the generalization of Tsallis and Renyi measures leads to Sharma-Mittal
information describing its formal structure in terms of KN-averages (for some further details see
\cite{Frank}, \cite{Masi}, \cite{Naudts}).

\subsection{How to construct generalized KN-averaged information-entropy measures}

The point of departure is the well known BG entropy \begin{equation}S_{BG}(P)= - k \, \sum_{i}
p_{i} \log p_{i} \, ,\end{equation} with $P = \{p_{1}, ...p_{\Omega}\}$ $(i=1,...,\Omega)$ a
probability distribution (PD) with $p_{i}$ the probability of the system to be in the i-th
microstate, $k$ the Boltzmann constant and $\Omega$ the total number of accessible microstates.
From now on we will use Shannon's information ($k=1$) and write it \begin{equation}S_{S}(P) =
\sum_{i} p_{i} \log \left( \frac{1}{p_{i}} \right) \, .\label{shannon} \end{equation} Given two
PD's $A$ and $B$, BG and Shannon's entropies follow an additive law \begin{equation} S_{S}(A \cap
B) = S_{S}(A) + S_{S}(B|A) \, , \label{shankin}\end{equation} with $S_{S}(B|A) = \sum_{i} \,
p_{i}(A) \, S_{S}(B|A_{i})$ the conditional information of B given A. Shannon's measure is also
recovered if one imposes axiomatically additivity through Shannon-Khinchin's axioms
\cite{Khinchin}.

However, additivity and/or extensivity can be violated in systems governed by nonlinear dynamics,
with long range interaction forces, with fractal structure or for other "non-ordinary"
deterministic types of behavior. According to a research trend that attempts to generalize
thermostatistics, these systems (but not all, see e.g. \cite{Chavanis}) can be better described by
a generalization of the BG-Shannon measure. This was first introduced by Havrda and Charvat
\cite{Havrda} in the information theory frame, and later by Tsallis for a generalized
thermostatistics \cite{Tsallis}: \begin{equation} S_{T}(P,q) = \frac{\sum_{i}p_{i}^{q}-1}{1-q} \,
= \, \frac{1}{q-1} \sum_{i}p_{i} \, (1- p_{i}^{q-1}) \, ,\label{tsallis} \end{equation} with $q>0$
a real parameter and where for $q\rightarrow1$ one recovers Shannon's measure. It can be written
also as a \textit{q-deformed Shannon entropy}\begin{equation} S_{T}(P,q) = - \sum_{i}p_{i}^{q}
\log_{q}p_{i} = \sum_{i} p_{i} \log_{q}\left(\frac{1}{p_{i}}\right) \,
,\label{q-shannon}\end{equation} where the \textit{generalized q-logarithm function}
\begin{equation} \log_{q} x = \frac{x^{1-q}-1}{1-q} \, , \label{q-log} \end{equation} has been
introduced, and which, for $q=1$, becomes the common natural logarithm.

Having introduced power law scaled PD's, Beck and Schl\"{o}gl defined \cite{Schloegl} what they
called \textit{escort-distributions} \begin{equation} \pi_{i}(P,q) = \frac{p_{i}^{q}}{\sum_{j}
p_{j}^{q}} \, ; \hspace{10mm} q>0 , \,\,\, q \in \mathbb{R} \, .\label{rho} \end{equation}
Accordingly one defines as \textit{normalized q-average} or \textit{escort average} of an
observable $\mathcal{O}$
\begin{equation}\left< \mathcal{O} \right>^{p} = \sum_{i}\pi_{i}(P,q) \, \mathcal{O}_{i} \,.\end{equation} Tsallis entropy
is \textit{pseudo-additive} \begin{equation}S_{T}(A \cap B) = S_{T}(A) + S_{T}(B|A) +
(1-q)S_{T}(A) S_{T}(B|A) \, .\label{tadditive} \end{equation} where
\begin{equation}S_{T}(B|A) = \sum_{i}\pi_{i}(A) \, S_{T}(B|A=A_{i}) \, ,\end{equation} with $\pi_{i}(A)$ defined in
\eqref{rho}.

In a similar fashion as we have seen for \eqref{shankin}, the generalization to conditional
entropies is possible too. S. Abe \cite{Abe} showed that the Shannon-Khinchin axioms can be
generalized to Tsallis' entropy introducing what he called the \textit{non-extensive conditional
entropy} $S_{T}(B|A)$.

Writing entropies with the inverse of the PD, like in the r.h.s. of \eqref{shannon} and
\eqref{q-shannon}, highlights its structure better since in a generalized context
$\log_{s}x^{\alpha} \neq \alpha \log_{s}x$ if $s\neq 1$. Then one sees that an information measure
is an average of the \textit{elementary information gains} \begin{equation}I_{i} \equiv
I_{i}\left(\frac{1}{p_{i}}\right) = \log_{s}\left(\frac{1}{p_{i}}\right)\end{equation} associated
to the i-th event of probability $p_{i}$, which for Shannon's and Tsallis' entropy means
\begin{equation} S_{S}(P) = \left< \log \left(\frac{1}{p_{i}}\right) \right>_{lin}
\label{shannon2} \end{equation} and \begin{equation} S_{T}(P) = \left< \log_{q}
\left(\frac{1}{p_{i}}\right) \right>_{lin} \label{tsallis2} \end{equation}

respectively, with the common \textit{arithmetic-}, or \textit{linear mean} $S = \sum_{i} p_{i}
I_{i} $.

A.N. Kolmogorov and M. Nagumo (\cite{Kolmogorov2}, \cite{Nagumo}) realized independently from each
other that, if Kolmogorov's axioms are supposed to be the foundation of probability theory, then
the notion of average can be generalized in more abstract terms to what is called a
\textit{quasi-arithmetic} or \textit{quasi-linear mean}. It turns out to be
\begin{equation} S = f^{-1} \left( \sum_{i} p_{i} \, f(I_{i}) \right) \, , \label{infomean}
\end{equation} with $f$ a strictly monotonic continuous function defined on real numbers $I_{i} $, called the
\textit{Kolmogorov-Nagumo function} (KN-function) and $p_{i}$ $(i=1, ..., \Omega)$ non-negative
weighting constants such that $\sum_{i}p_{i}=1$.\footnote{Typical examples are: $f(x)=x$ the
weighted mean (arithmetic mean if every $p_{i}= 1/\Omega$), $f(x)=x^{2}$ the root mean square,
$f(x)=\log x$ the geometric mean, etc. Kolmogorov and Nagumo considered only the case of the
arithmetic means (equal weights), while the generalization to arbitrary weights as given in
\eqref{infomean} was introduced by B. de Finetti \cite{Finetti} and T. Kitagawa \cite{Kitagawa}
and others later.}

Later Renyi showed \cite{Renyi} that if one imposes additivity on an entropy measure then there
are only two possible KN-functions left. The first leads to the conventional linear mean with
KN-function \begin{equation}f(x) = x \, ,\end{equation} and the second is the \textit{exponential
mean} represented by KN-function \begin{equation}f(x) = c_{1} \, b^{(1-q)x} + c_{2} \, ;
\hspace{10mm} q \in \mathbb{R}\label{quasilin} \end{equation} with $c_{1}$ and $c_{2}$ two
arbitrary constants and $b$ the logarithm base (we set $b$=e).

Shannon entropy comes from the former KN-function averaging over the elementary information gains
$I_{i} = \log (1/p_{i})$. \textit{Renyi's information measure} uses the latter one but averages
the same information gains ``exponentially'' and writes \begin{equation} S_{R}(P,q) =
\frac{1}{1-q} \log \sum_{i} p_{i}^{q} \, . \label{Renyi} \end{equation} For $q\rightarrow1$ it
becomes Shannon entropy. Renyi in this sense extended information measures but maintained it to be
additive.

Now, $S_{R}(P,q)$ is invariant under the choice of $c_{1}$ and $c_{2}$. Therefore we don't loose
in generality if in \eqref{quasilin} one sets $c_{1}=\frac{1}{1-q}=-c_{2}$, which, because of
\eqref{q-log}, is equivalent to \begin{equation}f(x) = log_{q} \, e^{x} \, ,\label{Renyifun}
\end{equation} and inserted in \eqref{infomean} with \begin{equation}I_{i}=\log \left(\frac{1}{p_{i}}\right) \, ,\end{equation}
shows that \eqref{Renyi} is \begin{equation}S_{R}(P,q) = \left< \log \left( \frac{1}{p_{i}}
\right) \right>_{\!\!\mathrm{exp}} \, ,\end{equation} with $\left<\cdot\right>_{exp}$ an average
defined by KN-function \eqref{Renyifun}. Compare this with \eqref{shannon2} and \eqref{tsallis2}.
One might call Renyi's entropy also an ``exponentially averaged Shannon entropy''. Napier's
logarithm assures additivity while \eqref{Renyifun} imposes the exponential average.

It should be noted how P. Jizba and T.Arimitsu \cite{jaRenyi} showed that Renyi's measure can be
obtained also from the extension of the Shannon-Khinchin axioms \cite{Khinchin} to a quasi-linear
conditional information \begin{equation}S_{R}(B|A)=f^{-1}\left( \sum_{i} \pi_{i}(A) f
\left(S_{R}(B|A_{i})\right) \right)\, , \end{equation} with $f$ as given in \eqref{quasilin}.

One might ask if there exists a measure which is non-additive and non-linearly averaged but
contains Tsallis' and Renyi's entropies as special cases. There are two ways to do this. The first
by working with the generalized exponentials and logarithms, the other through an extension of the
KN-mean.

Note that Tsallis and Renyi entropies have a special relationship: \begin{equation} S_{R}(P,q)
=\frac{1}{1-q} \log \left[ 1+ (1-q) \, S_{T}(P,q) \right] \, .\end{equation} By using generalized
logarithms \eqref{q-log} and the generalized exponential \begin{equation}e_{q}^{x} \equiv
e_{q}[x]=[1+(1-q)x]^{\frac{1}{1-q}} \, ,\end{equation} (which becomes the exponential function for
$q=1$) this becomes \begin{equation}S_{R}(P,q) = \log e_{q}^{S_{T}(P,\,q)} \, ,\label{ireqt}
\end{equation} and obviously \begin{equation}S_{T}(P,q) = \log_{q} e^{S_{R}(P,\,q)} \,
.\label{itqir}
\end{equation}

Eq. \eqref{ireqt} suggest immediately a further two parametric generalization:
\begin{equation}S_{SM}(P,\{q,r\})=\log_{r} e_{q}^{S_{T}(P,\,q)} = \frac{1}{1-r} \left[ \left(
\sum_{i} p_{i}^{\,q} \right) ^{\frac{1-r}{1-q}} - 1 \right] \, , \label{eqit}
\end{equation}

with $r$ another real parameter.\footnote{From \eqref{itqir} one naturally obtains also another
measure: $S_{SE}(P,\{q,r\})= \log_{q}e_{r}^{S_{R}(P,\,q)} = \frac{\left[ 1 + \frac{(1-r)}{(1-q)}
\, \log \sum_{i}p_{i}^{q}\right]^{\frac{1-q}{1-r}}-1}{1-q}$. We called it "supra-extensive" in
\cite{Masi}. It does however no longer follow a KN averaging theory and therefore we will not
consider it here.}

Equation \eqref{eqit} turns out to be B.D. Sharma and D.P. Mittal's information measure
\cite{Sharma}. They used it in the frame of information theory about thirty years ago and it was
more recently investigated in the context of a generalized thermostatistics by Frank,
Daffertshofer and Plastino (\cite{Frank}, \cite{Daffertshofer}).

If one remembers how the q-logarithm satisfies the pseudo-additive law \begin{equation}\log_{q} xy
= \log_{q} x + \log_{q} y + (1-q) (\log_{q} x) (\log_{q} y) \end{equation} and notes the
decomposition
\begin{equation}e_{q}^{x + y + (1-q)x y } = e_{q}^{x} \, e_{q}^{y} \, ,\end{equation} then, in this representation, it is
immediately visible that a pseudo-additive law is satisfied also for SM entropy (start from
\eqref{eqit}, use \eqref{tadditive}) \begin{equation}S_{SM}(A \cap B) = \log_{r}e_{q}^{S_{T}(A
\cap B)} = \log_{r}(e_{q}^{S_{T}(A)+S_{T}(B|A)+(1-q)S_{T}(A)S_{T}(B|A)})=\end{equation}
\begin{equation} = \log_{r}(e_{q}^{S_{T}(A)} e_{q}^{S_{T}(B|A)}) = S_{SM}(A) + S_{SM}(B|A) +
(1-r)S_{SM}(A) S_{SM}(B|A) \, ,\end{equation} with $A$ and $B$ the families of PDs of two
statistical systems.

Note that it is the magnitude of parameter $r$ which stands for the degree of pseudo-additivity,
while $q$ stands for a PD deformation parameter. In Tsallis' entropy, when $r \rightarrow q$, the
pseudo-additive parameter $r$ merges into the deformation parameter $q$, and outside the frame of
a KNG-representation, the two appear somehow mixed together.

But, and this is the equivalent second method, SM entropy emerges also by q-generalizing in the
most simple possible way the KN-mean \eqref{Renyifun} to \begin{equation}f(x) = log_{q} \,
e_{r}^{x} \, ,\end{equation} which we call the \textit{quasi-exponential mean}, and \eqref{eqit}
can be written as
\begin{equation}S_{SM}(P,\{q,r\}) = \left<\log_{r}\left(\frac{1}{p_{i}}\right) \right>_{\!\!\mathrm{q-exp}} \,
.\end{equation}

\subsection{The generalized information-entropy measures hierarchy}

Manipulating with the formalism, frequently it turns out useful to rewrite in different
alternative ways the quantity \begin{equation}\left(\sum_{i}p_{i}^{q}\right)^{\frac{1}{1-q}} \!\!
= \left< \left(\frac{1}{p_{i}}\right)^{1-q}\right>^{\frac{1}{1-q}}_{\!\mathrm{lin}} \!\! =
e_{q}^{\left<log_{q}\left(\frac{1}{p_{i}}\right)\right>_{\!\mathrm{lin}}}=e_{q}^{S_{T}(P,q)}
=e^{S_{R}(P,q)} \, . \label{eii}\end{equation}

But writing it in the KNG-representation leads also to the following equality
\begin{equation}\Gamma(P,q)\equiv \left<\frac{1}{p_{i}}\right>_{\!\!\log_{q}} \!\!\!\! =
\left(\sum_{i}p_{i}^{q}\right)^{\frac{1}{1-q}} \!\! \, ,\label{eii2}\end{equation} where we used
what we call the \textit{generalized logarithmic mean} $\left< \cdot \, \right>_{log_{q}}$ defined
by the KN-function $f(x)=log_{q}x$.

Then, entropies \eqref{ireqt}, \eqref{itqir} and \eqref{eqit}, using \eqref{eii} in form of
\eqref{eii2}, can be written in an overall generalized hierarchy of information measures as
follows

\begin{equation}\hspace{-0mm} S_{SM}(P,\{q,r\}) = \log_{r}
\Gamma(P,q) =\, \log_{r} \left<\frac{1}{p_{i}}\right>_{\!\!\log_{q}} \,
;\label{nonrelnaudts}\end{equation}

\begin{equation}\hspace{-0mm} S_{T}(P,q)  = \log_{q}
\Gamma(P,q)=\, \log_{q} \left<\frac{1}{p_{i}}\right>_{\!\!\log_{q}} \,
;\label{nonreltsallis}\end{equation}

\begin{equation} \hspace{-0mm} S_{R}(P,q) = \log \Gamma(P,q)=\,  \log \left<\frac{1}{p_{i}}\right>_{\!\!\log_{q}} \,
;\label{nonrelrenyi}
\end{equation}

\begin{equation}S_{S}(P) = \log \Gamma(P,1) =\, \log \left<\frac{1}{p_{i}}\right>_{\!\!\log}\, ,\label{nonrelshannon}\end{equation}

where also we added Shannon's measure.

Then one sees how SM's entropy generalizes Renyi's additive entropy to pseudo-additivity,
characterized by the r-logarithm. This formalism makes it also easier to recognize the behavior of
the limits for $r \rightarrow q$ or $r \rightarrow 1$ of \eqref{nonrelnaudts} than in their
explicit form (the r.h.s. of \eqref{eqit}). Looking also at its "$\log_{r} \times \exp_{q}$
\textit{form}" (the middle term of \eqref{eqit}) it is immediately visible (without any need to
apply Hopital rule, first order approximations or whatever) how, for $r \rightarrow q$ and $r
\rightarrow 1$, they reduce to Tsallis and Renyi entropy respectively.

According to ordinary extensive statistical mechanics entropy can be defined as $S=k \log\Gamma$,
with $\Gamma$ the entire phase-space volume of the system ($\Gamma$-space). In the
KNG-representation, from \eqref{nonrelshannon} (and unitary Boltzmann constant) this is given by
$\Gamma(P,1) =\, \log \left<\frac{1}{p_{i}}\right>_{\!\!\log} =
e^{\,S_{S}(P,1)}=e^{\sum_{i}p_{i}\log\frac{1}{p_{i}}} \, .$ While in general, the above
 information-entropy measures hierarchy is highlighted in terms of KN q-generalized $\Gamma$-space
volumes $\Gamma(P,q) = \left<\frac{1}{p_{i}}\right>_{\!\!\log_{q}} =
\left(\sum_{i}p_{i}^{q}\right)^{\frac{1}{1-q}}$.

\subsection{Some remarks on the generalizing path taken}

One can see \cite{Frank} that MEPDs (which from now on will be labeled with a caret symbol
$\widehat{}$ ) obtained from SM entropy subjected to the escort Hamiltonian
\begin{equation}\left<H\right>  = \frac{\sum_{i}p_{i}^{q}E_{i}}{\sum_{i}p_{i}^{q}} = U_{g}
\label{pichamil}\end{equation} are of the form \begin{equation}\widehat{p}_{i}^{\,SM} =
\frac{e_{q}[-\beta_{g}(E_{i}-U_{g})]}{Z_{SM}} =
\frac{e_{q}[-\overline{\beta}_{g}E_{i}]}{\overline{Z}_{SM}} \label{pitotal} \end{equation} with
\begin{equation}Z_{SM} = \sum_{j} e_{q} \left[-\beta_{g} (E_{j}-U_{g}) \right] \, ;
\hspace{5mm} \overline{Z}_{SM} = \sum_{j} e_{q} \left[-\overline{\beta}_{g} E_{j}\right] =
\frac{Z_{SM}}{e_{q}[\beta_{g}U_{g}]}  \label{pitotal2} \end{equation} the generalized partition
function, $\beta_{g}$ the generalized Lagrangian parameter (where the parameter $r$ has been
absorbed) and $E_{j}$ the microstate energy.\footnote{Note that in general eq. \ref{pitotal} does
not maintain itself always positive and real without cut-off prescriptions. More on this can be
found in \cite{Masi3} and references therein.} The terms on the r.h.s. of \eqref{pitotal} and
\eqref{pitotal2} highlight the mean energy translational invariance of the MEPDs with
$\overline{\beta}_{g} = \frac{\beta_{g}}{1 + (1-q) \beta_{g}U}$.

It is therefore important to remember that a q-generalized Boltzmann factor does not necessarily
indicate that one is dealing with Tsallis statistics. It is a more general feature which is common
to Renyi and SM entropic functionals too. It is instead $\beta_{g}$ (or $\overline{\beta}_{g}$)
which selects one or the other statistics. In this regards let us shortly comment on an issue
which needs some clarification when generalized entropies other than that of Tsallis are
considered in a generalized thermostatistics.

It should be mentioned that there are some reasons to argue that the Renyi entropy is not
Lesche-stable (for $q\neq1$), is concave only for $q\leq1$, and does not possess the property of
finite entropy production per unit time. And one should expect that further generalizations of it,
like SM entropy, cannot possess these properties either.\footnote{But one should also be aware of
the fact that, at least in principle, it still can recover it for a limited r-parameter space with
$r\neq1$.} If this has some thermodynamical implications is currently a controversial matter.
There have been different types of critiques to Lesche's and Abe's approach (see e.g. some papers
listed in \cite{controversy}). It seems to us that this is still an open issue that so far has not
been clarified entirely.

Moreover, the kind of statistics which emerges here is two parametric. While so far no evidence
exists that natural systems are governed by more than a one parameter extended BG statistics, we
will try to put forward some arguments which indicate that an extra parameter might furnish some
new insights on the difficulties related to the contrasting physical interpretations and
microscopic justification of the $q$-parameter (e.g. \cite{Almeida}, \cite{Chavanis},
\cite{Beck}). And fact is also that from an analytic point of view, as Frank and Plastino showed
\cite{Frank}, neither Renyi nor Tsallis entropy are the most general pseudo-additive measures
which allow for escort averages admitting of a partition function, but SM's measure is, which is
two parametric. SM entropy is also consistent with the familiar q-invariant Legendre transform
structure of thermodynamics and the generalizations here considered emerge in any case as a
natural consequence of Kolmogorov's axiomatics applied to power law PDs.

Anyway, even if not in the frame of an extended thermostatistics, these measures still can have an
application at least in other research fields of information theory, as cybernetics, control
theory, signal processing, cryptography or in a large class of chaotic dynamical system or in
other areas of mathematical statistics.

\section{The meaning of the $q \rightarrow \frac{1}{q}$ duality in the frame of the KNG-representation and some possible consequences\label{secdual}}

The notion of duality is used in rather different connections. Generally speaking, two physical
theories are said to be dual if there exists a map, a duality transformation, so that the first
theory can be transformed into a second theory looking just like the first one. A duality acts by
interchanging the roles of two objects linking quantities that seemed to be separate (e.g. the
electric with the magnetic fields, other strong and weak coupling strengths, large and small
distance scales or, as in our case, two PDs which dually link the expectation values of an
observable, like that of a Hamiltonian). These quantities have always been distinctive of some
limits of a physical system in both classical and quantum field theories and it is then possible
to observe the behavior of the observable in one model in terms of the other and eventually use
perturbative and semiclassical methods.

Recent studies highlighted how the Tsallis entropy seems to lead just naturally to a dual
description in terms of a $q \leftrightarrow \frac{1}{q}$ duality too. Its function consists in
linking the distribution $P = \{ p_{i}...p_{\Omega}\}$ with the escort distribution $\Pi = \{
\pi_{i}...\pi_{\Omega}\}$ of \eqref{rho}, that is \begin{equation} \pi_{i}(P,q) =
\frac{p_{i}^{q}}{\sum_{j}p_{j}^{q}} \hspace{5mm} \leftrightarrow \hspace{5mm} p_{i}(\Pi,q) =
\frac{\pi_{i}^{1/q}}{\sum_{j}\pi_{j}^{1/q}} \, .\label{pipai}\end{equation} If one considers $\Pi$
and $P$ both normalized distributions it is easy to verify this by simply inserting one identity
in the other.

This duality was first noted by Tsallis, Mendes and Plastino \cite{Tsallis4}, and further
investigated by Frank and Plastino \cite{Frank}, Naudts \cite{Naudts2}, and others (even earlier,
Abe noted a similar duality \cite{Abe2}, however of a quite different nature). We will call it the
Tsallis, Mendes and Plastino duality, or simply the TMP-duality.

This dually relates also the q-expectation of an observable $\mathcal{O}$ with its "classical"
average as\footnote{The fact that the TMP-duality maps an escort averaged observable into a
linearly averaged one should not be confused with a transformation of a q-statistics into a BG
one! The PD involved in this linear mean are here defined in the frame of a generalized
statistics.}
\begin{equation}\left<\mathcal{O}\right>^{p}=\frac{\sum_{i}p_{i}^{\,q}\mathcal{O}_{i}}{\sum_{j}p_{j}^{q}}
\,\,\,\, \leftrightarrows \,\,\,\,
\frac{\sum_{i}\pi_{i}^{1/q}\mathcal{O}_{i}}{\sum_{j}\pi_{j}^{1/q}} =
\left<\mathcal{O}\right>^{\pi} \equiv \left<\mathcal{O}\,\right>\,.\end{equation} Here we will try
to understand its deeper meaning and underline how the TMP-duality does not characterize only
Tsallis' entropy but is a general transformation which works for every measure in a q-generalized
statistics.

\subsection{The TMP-duality as KNG-representation form invariant transformation \label{secone}}

The TMP-duality has a particularly important effect in transforming the phase-space volume which
becomes clear in the KNG-representation. Transformation \eqref{pipai} leads to the following
identity \begin{equation} \Gamma(P,q) \equiv \left<\frac{1}{p_{i}}\right>_{\!\!\log_{q}} =
\left(\sum_{i}p_{i}^{q}\right)^{\frac{1}{1-q}} = \end{equation}
\begin{equation} = \left( \sum_{i} \pi_{i}^{1/q} \right)^{\frac{-q}{1-q}}=
\left<\frac{1}{\pi_{i}}\right>_{\!\!\log_{1/q}} \equiv \Gamma(\Pi,\frac{1}{q}) \,
.\label{degpipai}\end{equation}

Simply insert the expression for $p_{i}$ of \eqref{pipai} remembering the normalization condition
and evaluate the logarithmic mean of $\frac{1}{\pi_{i}}$ with $q \rightarrow \frac{1}{q}$
according to \eqref{eii2}.

Therefore the information-entropy hierarchy structure of \eqref{nonrelnaudts} to
\eqref{nonrelshannon} transforms into \begin{equation}\hspace{-0mm} S_{SM}(\Pi,\{q,r\}) = \log_{r}
\Gamma(\Pi,\!\frac{1}{q}) = \log_{r} \left<\frac{1}{\pi_{i}}\right>_{\!\!\log_{1/q}}  \,
;\label{nonrelnaudts2}\end{equation}

\begin{equation}\hspace{-0mm} S_{T}(\Pi,q)  = \log_{q}
\Gamma(\Pi,\!\frac{1}{q}) = \log_{q} \left<\frac{1}{\pi_{i}}\right>_{\!\!\log_{1/q}} \,
;\label{nonreltsallis2}\end{equation}

\begin{equation} \hspace{-0mm} S_{R}(\Pi,q) = \log \Gamma(\Pi,\!\frac{1}{q}) = \log \left<\frac{1}{\pi_{i}}\right>_{\!\!\log_{1/q}}\,
;\label{nonrelrenyi2}
\end{equation}

\begin{equation}
S_{S}(\Pi) = \log \Gamma(\Pi,\!1) = \log \left<\frac{1}{\pi_{i}}\right>_{\!\!\log}\,
.\label{nonrelshannon2}
\end{equation}

Eqs. \eqref{nonrelnaudts2} to \eqref{nonrelshannon2} represent the same entropy measures
\eqref{nonrelnaudts} to \eqref{nonrelshannon} after the transformation \eqref{pipai}. One does
accordingly speak of a $p$\textit{-picture} and a $\pi$\textit{-picture statistics}. This makes it
clear how a dual description between a $q>1$ and a $q<1$ statistics exists. By means of the
KNG-representation, and once identity \eqref{degpipai} is known, the TMP-duality between the
$p$-picture and the $\pi$-picture statistics follows immediately without need for further proof.

In the frame of the KNG-representation the TMP-duality reveals to be a "measure form invariant"
transformation, i.e. it maintains the q-logarithmic mean over the inverse PDs. A different duality
than that of \eqref{pipai} would not lead to same invariance. For instance another duality that
has received some attention in the literature is \begin{equation}\psi: q \leftrightarrow 2-q \,
.\end{equation}

One can accordingly speak of a \textit{$\psi$-picture} in this case. We might ask therefore if
also this transformation shares similar analytic properties. However, in this case there seems to
be no way to define a correspondent dual redefinition of the PD in the $\psi$-picture which
describes a normalized distribution and that at the same time leads to form-invariance of the
measures, as it is in the case with the TMP-duality. In fact this would imply that, in the same
fashion of \eqref{degpipai}, there must exist a PD $\Psi= \{\psi_{1},...,\psi_{\Omega}\}$ such
that
\begin{equation} \Gamma(P,q) = \left<\frac{1}{p_{i}}\right>_{\!\!\log_{q}} =
\left(\sum_{i}p_{i}^{q}\right)^{\frac{1}{1-q}} = \left( \sum_{i} \psi_{i}^{2-q}
\right)^{\frac{1}{q-1}}= \left<\frac{1}{\psi_{i}}\right>_{\!\!\log_{2-q}} \,
,\label{pipai2}\end{equation} and satisfying the normalization condition so that it must have a
structure like
\begin{equation}\psi_{i}= \frac{g_{i}(p_{i},q)}{\sum_{j}g_{j}(p_{j},q)} \, ,\end{equation} with
$g_{i}$ some map that transforms smoothly every PD $p_{i}$ of the $p$-picture into every
$\psi_{i}$ of the $\psi$-picture and such that for $q \rightarrow 1$ it goes like $g_{i}(p_{i},q)
\rightarrow p_{i}$. Inserted in the second r.h.s. term of \eqref{pipai2} this means that we are
searching for a $g_{i}(p_{i},q)$ that satisfies to
\begin{equation}\frac{\left(\sum_{i}g_{i}(p_{i},q)\right)^{\,2-q}}{\sum_{j} g_{j}(p_{j},q)^{2-q}}
= \sum_{k}p_{k}^{q} \, .\end{equation} This seems not to have a simple general solution.

The TMP-duality is special in the sense that, given a normalized PD, it preserves the associated
generalized information-entropy measures KNG-representation form invariant.

Note that in Tsallis' measure \eqref{nonreltsallis2} one transforms the deformation parameter $q
\rightarrow \frac{1}{q}$ of $\Gamma$, but not the non-additive parameter $r=q$. This makes sense
also because otherwise the entropy scalar would not be invariant ($\log_{q} \Gamma(P,q) = \log_{q}
\Gamma(\Pi,\!\frac{1}{q}) \neq \log_{\frac{1}{q}} \Gamma(\Pi,\!\frac{1}{q})$) and the entropy in
the $\pi$-picture would no longer represent the measure of canonical phase-space volume, against
the statistical definition of entropy itself. The two $q$-parameters of Tsallis' entropy
\eqref{tsallis} should be handled separately. This might suggest to distinguish between a
pseudo-additive and non-extensive parameter. Indeed there is a subtle but potentially important
distinction between these two notions which is frequently overlooked (see e.g. \cite{Touchette}
and \cite{Tsallis3}). A single parameter statistics which tries to intertwine them together might
lead to an informational loss.

\subsection{The TMP-duality as a non-extensive Kramers-Wannier duality\label{sectwo}}

There is a known relation which connects the partition function with the $\Gamma$-space of a
stationary system. One can generalize to an arbitrary $(\widehat{\varepsilon}, q\,')$-picture
setting ($\widehat{\varepsilon} = \widehat{p}$, $q'=q$) for the $p$-picture,
($\widehat{\varepsilon} = \widehat{\pi}$, $q'=\frac{1}{q}$) for the $\pi$-picture, and label
$\widehat{P}$ and $\widehat{\Pi}$ the corresponding MEPD's family. Then, writing the mean energies
as $U_{g}^{p} = <\!\!H\!\!>^{p}$ and $U_{g}^{\pi} = <\!\!H\!\!>^{\pi} =<\!\!H\!\!>$, we have
\begin{equation} Z_{SM}(\widehat{\varepsilon}) = \sum_{j} e_{q\,'} \left[-\beta_{g}^{\varepsilon}
(E_{j}-U_{g}^{\varepsilon}) \right] =
\left(\sum_{i}\left(\widehat{\varepsilon}^{\,SM}_{i}\right)^{q\,'}\right)^{\frac{1}{1-q\,'}}
\!\!\!\!= \left<\frac{1}{\widehat{\varepsilon}^{\,SM}_{i}}\right>_{\log_{q\,'}} \!\!\!\!\! = \,
\Gamma(\widehat{\varepsilon},q\,') \, .\label{deltapart}\end{equation} Because, from
\eqref{pitotal} we have
$$\left(\widehat{\varepsilon}^{\,SM}_{i}\right)^{\,1-q\,'} \left(Z_{SM}(\widehat{\varepsilon})\right)^{1-q\,'} = \left(e_{q\,'}\left[-\beta_{g}^{\varepsilon}(E_{i}-U_{g}^{\,\varepsilon})\right]\right)^{1-q\,'} = 1 -(1-q\,') \beta_{g}^{\varepsilon}(E_{i}-U_{g}^{\,\varepsilon}) \,.$$ Then multiplying both members with $\left(\widehat{\varepsilon}^{\,SM}_{i}\right)^{\,q\,'}$ and summing up, we have
$$\sum_{i} \widehat{\varepsilon}^{\,SM}_{i} \left(Z_{SM}(\widehat{\varepsilon})\right)^{1-q\,'} =
\left(Z_{SM}(\widehat{\varepsilon})\right)^{1-q\,'} =$$ $$ =
\sum_{i}\left(\widehat{\varepsilon}^{\,SM}_{i}\right)^{q'} - (1-q\,')
\beta_{g}^{\varepsilon}\sum_{i}\left(\widehat{\varepsilon}^{\,SM}_{i}\right)^{q\,'}(E_{i}-U_{g}^{\,\varepsilon})=
\sum_{i}\left(\widehat{\varepsilon}^{\,SM}_{i}\right)^{\,q\,'} \, ,$$ because of the normalization
condition and where in the last passage the constraint \eqref{pichamil} has been imposed. From
this one obtains \eqref{deltapart}.

And since $Z_{SM}(\widehat{P})= \Gamma(\widehat{P},q)$ and $Z_{SM}(\widehat{\Pi}) =
\Gamma(\widehat{\Pi},\frac{1}{q})$, then because of \eqref{degpipai}, what the TMP-duality really
implies is that \begin{equation}Z_{SM}(\widehat{P})= Z_{SM}(\widehat{\Pi}) \, .\end{equation} The
partition function represents something which is "picture invariant"\footnote{One has not to
accept SM-entropy as a measure for thermostatistics to reach the same conclusion because this
holds also for Tsallis entropy once we identify \eqref{eii2} with a q-generalized phase-space
volume. In the present context we have shown however that this is a more general feature of the
formal structure which stands behind Renyi and SM-entropies too.}.

On the other side, always because of \eqref{pitotal2}, this means also that
\begin{equation}\sum_{j} e_{q} \left[-\overline{\beta}_{g}^{\,p} E_{j} \right] = \sum_{j} e_{1/q}
\left[-\overline{\beta}_{g}^{\,\pi} E_{j}\right] \, , \end{equation} which in general can't hold
for the same generalized Lagrangian parameters (i.e. for $\overline{\beta}_{g}^{\,p}=
\overline{\beta}_{g}^{\,\pi}$ and $q\neq1$). The TMP-duality establishes therefore a
correspondence between the partition functions of two equilibrated systems with different PDs
determined by a Hamiltonian with a $q>1$ and a $q'=\frac{1}{q}<1$ statistics and by the Lagrangian
parameter $\overline{\beta}_{g}^{\,p}$ and $\overline{\beta}_{g}^{\,\pi}$ respectively.

This equality between the generalized partition functions at two different
temperatures\footnote{What kind of relation exists between the generalized non-extensive
Lagrangian parameter and the ordinary statistical notion of "temperature" is actually a matter of
debate, and what kind of non-trival rescaling of the system's temperatures in the different
pictures must be employed is therefore not entirely clear (see however a paper by Salazar and
Toral \cite{Salazar2} in this sense), but this is at present not so decisive for the present work
and we will generically speak of a system at temperature $T^{p}$ and $T^{\pi}$ for the respective
pictures.} is the q-generalized analogue of the Kramers-Wannier two-dimensional Ising model
duality \cite{Kramers} between an ordered and disordered phase (see also \cite{Wegner}, and for a
more modern introduction \cite{Kadanoff}). In this scenario a connection is established between a
randomly ordered system (typically a non magnetized ferromagnetic substance) above the critical
Curie temperature with another system at less than the critical temperature undergoing a symmetry
breaking (typically represented by the formation of domain walls).

There have been some quite interesting attempts to expand the Ising model to non-extensive
regimes. It can be taken as a good starting point to investigate non-extensivity of systems with
long-range interactions presenting critical properties (for some pioneering works in this field
see e.g. \cite{Salazar2}, \cite{Andrade}, \cite{Andrade2},  \cite{Nobre}, \cite{Jund},
\cite{Salazar}, \cite{Cannas},  \cite{Botet}, \cite{Andrade3}, and others listed in
\cite{Tsallis2}).

More precisely, one can extend the Ising ferromagnet to a long-range interaction Hamiltonian
(without magnetic interactions) as \begin{equation}H\{\sigma_{i}\} = -
\sum_{<ij>}\!\!J(\!J_{0},\alpha) \, \sigma_{i}\,\sigma_{j} \hspace{10mm} (\sigma_{i}= \pm 1
\forall i) \end{equation} with
\begin{equation}J\,(\!J_{0},\alpha) = \frac{J_{0}}{r_{ij}^{\,\alpha}} \hspace{10mm} (J_{0}>0; \, \alpha >0) \, ,\end{equation}
where $r_{ij}$ is the distance (in crystal units) between the i-th and j-th site, $\sum_{<ij>}$ is
a sum running over all distinct sites of pairs, $J_{0}$ is the exchange coupling constant and
$\alpha$ determines the power law of the long-range interaction. For $\alpha \rightarrow \infty$
one recovers the first-neighbor model, while for $\alpha \rightarrow 0$ (and rescaling $J_{0}
\rightarrow J_{0}/N$) the Curie-Weiss model.

The overall picture which emerges from the above mentioned investigations is that this long-range
spin model indeed exhibits non-extensive scaling laws and phase transitions at critical
temperatures which depend on $\alpha$ and the space dimensionality $d$ of the system. In
particular it turns out that as long as $\alpha > d$ the system is extensive, whereas for $0\leq
\alpha \leq d$ the behavior is non-extensive. The $q$-parameter seems to depend exclusively on the
dimensionality and the long-range interaction power law, i.e. $q\equiv q(\alpha,d)$.

Therefore for the 2d-Ising model the partition function in the $(\varepsilon,
q'(\alpha^{\varepsilon}))$-picture can be written
\begin{equation}Z_{SM}(\varepsilon) = \sum_{\sigma_{1}} \,
\sum_{\sigma_{2}}...\sum_{\sigma_{N}}
e_{q(\alpha^{\varepsilon})}[-\overline{\beta}_{g}^{\,\,\varepsilon} H\{\sigma_{i}\}] \,
,\end{equation} with $\alpha^{\varepsilon}$ the power-law parameter for the chosen picture.

The TMP-duality implies \begin{equation}Z_{SM}(\overline{\beta}_{g}^{\,p}, \alpha^{p}) =
Z_{SM}(\overline{\beta}_{g}^{\,\pi} , \alpha^{\pi}) \, , \label{dualz} \end{equation} i.e. it
dually relates a system at a temperature $T^{p}$ governed by a long-range interaction power law
$\propto 1/r^{\,\alpha_{p}}$ and another system at a temperature $T^{\pi}$ with interaction law
$\propto 1/r^{\,\alpha_{\pi}}$, \textit{and} with $\alpha^{\pi}$ subjected to the condition that
$q(\alpha^{\pi})= \frac{1}{q(\alpha^{p})}$.

\subsection{Some remarks on the conceptual foundations of the TMP-duality}

The probabilities transformation $p_{i} \rightarrow \pi_{i}$ in section \eqref{secone} can be seen
also as a "coordinate" transformation. Here one might compare the $\pi$-transformation which leads
to the KNG-representation form invariance as a statistical analogue of the Lorentz coordinate
transformations in special relativity. This analogy holds however only partially since the former
is a discrete and the latter a continuous transformation. The TMP-duality does not represent a
continuous physical transformation, like a "boost" in special relativity or, more specifically, an
isentropic (adiabatic and reversible) process. Moreover, in section \eqref{sectwo} it links dually
two systems at two different temperatures, but these are also governed by different interaction
forces among their units. Therefore this kind of statistical duality resembles more the
strong-weak coupling duality, acting not only on the coupling constants but also on the power laws
of the long-range forces determining the non-extensive behavior of the system. It relates two
problems with different couplings, mapping the strong coupling regime into the weak one. Dualities
seem to be an underlying common feature of physical theories in general, as for instance in the
case of the electromagnetic duality. Especially in modern quantum gravity theories, as in
superstring theories, these kind of dualities turned out to be useful in describing the theory
from a nonperturbative standpoint making calculations easier and uniting apparently different
theories in one comprehensive framework. The duality described here, i.e. that which leads to eq.
\eqref{dualz}, might potentially also represent a tool to simplify the calculations of the
thermodynamic properties of a complex non-extensive system governed by non-trival long-range
interaction force laws studying its (eventually extensive) dual counterpart with differently
scaled interaction forces. A more precise understanding of the conceptual foundations is of course
desirable.

\section{Conclusion}

Generalized information-entropy measures have been presented in the light of a formalism based in
particular on the manipulation of the generalized exponential and logarithm functions together
with Kolmogorov's and Nagumo's averaging. This kind of formal approach to a generalized statistics
seems to lend itself particularly well in describing the deeper implications of the $q
\leftrightarrow \frac{1}{q}$ duality of a q-generalized statistics which goes beyond that
described by Tsallis entropy functionals. We showed how this duality implies a KNG-representation
entropy form invariance. The $q \rightarrow 2-q$ duality seems to be less fundamental in this
sense: unlike the TMP-duality it does not preserve a KNG-representation form invariance for a
normalized PD. The TMP-duality establishes also an equivalence between two partition functions for
different Lagrangian parameters and long-range interaction power laws in the $p$ and
$\pi$-pictures respectively. A possible physical meaning of this is that it can be interpreted as
a 2d Ising model duality generalized to non-extensive systems.

\end{document}